\documentclass[doublecol]{epl2} 
\usepackage{amssymb}
\usepackage{color}
\title{Model Selection with Baryonic Acoustic Oscillations \\ in the Lyman-$\alpha$ Forest}
\shorttitle{BAO in the Lyman-$\alpha$ Forest} %Insert here a short version of the title if it exceeds 70 characters

\author{F. Melia%\inst{1}
}
\shortauthor{F. Melia}

\institute{                    
Department of Physics, The Applied Math Program, and Department of Astronomy, \\
The University of Arizona, AZ 85721, USA}
\pacs{98.80.Cq}{Inflation}
\pacs{98.80.-k}{Cosmology}
\pacs{98.80.Jk}{Relativistic Astrophysics}

\abstract{The recent release of the final, complete survey of Lyman-$\alpha$ baryonic
acoustic oscillation measurements provides the most significant and accurate data
base for studying cosmic geometry at an effective redshift $z_{\rm eff}=2.334$,
which is inaccessible to other sources. In this {\sl Letter}, we use these data to select
among four distinct cosmologies: Planck $\Lambda$CDM, the $R_{\rm h}=ct$ universe,
the Milne universe and Einstein-de Sitter. Given the breadth and depth of the
Lyman-$\alpha$ study, this BAO measurement alone provides a strong model comparison,
complementary to previous studies that combined Lyman-$\alpha$ data with measurements
at lower redshifts. Though both approaches are useful, the latter tends to dilute the
disparity between model predictions and the observations. We therefore examine how
the models compare to each other strictly based on the BAO scale measured in the
Lyman-$\alpha$ forest and background quasars. We find that Milne and Einstein-de
Sitter are strongly ruled out by these data. There is also strong evidence disfavoring 
the standard model. The Lyman-$\alpha$ measurements are completely consistent with 
the cosmic geometry predicted by $R_{\rm h}=ct$. As such, evidence continues to 
grow that the zero active mass condition from general relativity ought to be an
essential ingredient in $\Lambda$CDM.}

\begin{document}

\maketitle

\section{Introduction}\label{intro}
The angular power spectrum in the cosmic microwave background (CMB) is characterized by
an angular scale set at the surface of last scattering, when the radiation began 
dissociating from baryonic matter. Believed to represent a sonic horizon, $r_d$, 
this comoving distance is apparently associated with sound waves created at or near 
the Planck regime, which then propagated across the baryon-photon fluid until
the CMB was produced \cite{Peebles:1970,Sunyaev:1970,PlanckVI:2020}.  

In a Friedmann-Lema\^itre-Robertson-Walker (FLRW) cosmology, in which all proper (or physical) 
distances grow in proportion to a universal scale factor, $a(t)$, while comoving distances
remain constant, the scale $r_d$ is expected to have remained unchanged, reappearing 
subsequently as a characteristic length in the matter correlation function. In this context, 
the sonic scale is more commonly inferred from baryonic acoustic oscillations (BAO) seen
in the large-scale structure, the first measurement of which was carried out using the 
auto-correlation of galaxy positions \cite{Eisenstein:2005} at $z\sim 0.35$, and the 
galaxy power spectrum \cite{Cole:2005} at $z\sim 0.1$. More generally, the BAO scale 
at $z\lesssim 2$ has been studied using a variety of discrete tracers, such as 
galaxy clusters \cite{Hong:2016}, and quasars \cite{Ata:2018}, in addition to the
aforementioned galaxies \cite{Percival:2007,Blake:2011,Beutler:2011,Padmanabhan:2012,Anderson:2012,Ross:2015,Alam:2017,Bautista:2018}.

Beyond $z\sim 2$, however, a measurement of the BAO scale must to be handled differently
because the number density of observable dicrete tracers is too low for high precision 
clustering studies. The method of choice for measuring $r_d$ in the high-redshift 
Universe instead rests on the observation of opacity fluctuations in the Lyman-$\alpha$ 
forest irradiated by background quasars. The first such studies, focusing on the 
Lyman-$\alpha$ auto-correlation function, were carried out by refs.~\cite{Busca:2013}, 
\cite{Slosar:2013}, \cite{Kirkby:2013}, \cite{Delubac:2015}, \cite{Bautista:2017} 
and \cite{deSainte:2019}. Complementary results based on the Lyman-$\alpha$ and 
quasar cross-correlation function have also been reported by refs.~\cite{Font-Ribera:2014}, 
\cite{duMas:2017} and \cite{Blomqvist:2019}. 

The BAO feature is a powerful diagnostic in cosmology because it yields both 
angular-diameter distances and the expansion rate normalized to the sound horizon
$r_d$. And while the luminosity distance to type Ia supernovae has already provided
strong evidence for the existence of dark energy \cite{Riess:1998,Schmidt:1998,Perlmutter:1999},
these transient events are difficult to observe at redshifts $z\gtrsim 1.8$. 

The BAO measurements in the Lyman-$\alpha$ forest, at an effective $z\sim 2.334$, may 
therefore be used in several distinct tests of the geometry of the Universe at 
intermediate redshifts not accessible to local surveys focusing on supernovae ($z\lesssim
1.8$) and instruments designed to study the CMB at $z\gg 1$.

A common approach is to combine all of the BAO data, those from the galaxy surveys at low 
redshifts and those from the Lyman-$\alpha$ forest farther away, under the assumption that 
$r_d$ is independent of $z$. This can be done either in the context of $\Lambda$CDM, where 
one includes the predicted value of $r_d$ in this model to extract the redshift-dependent 
angular-diameter distance and expansion rate, or in model selection by examining which 
cosmology is preferred by the BAO measurements. For the latter, one considers
$r_d$ to be `unanchored,' since its value may not be the same from one model to the
next. In this case, one either optimizes $r_d$ along with the other model parameters
individually for each cosmology being tested, or avoids it altogether by considering 
ratios of the angular-diameter and Hubble distances, both of which are proportional 
to the sonic horizon (see Eqs.~\ref{eq:dHz} and \ref{eq:dMz} below).

In previous work \cite{MeliaLopez:2017,Melia:2020c,Melia:2022c}, we have followed the 
latter approach using older, less complete BAO catalogs than those available now to compare 
the standard model with one of its principal competitors known as the $R_{\rm h}=ct$ 
universe \cite{MeliaShevchuk:2012,Melia:2020}. Interestingly, the aggregated BAO
data have tended to favour the latter model rather than $\Lambda$CDM. But the BAO
measurements using the lower redshift galaxy surveys are less discerning than their 
higher redshift counterparts, so while the outcome of these studies has been
suggestive, it has not necessarily been compelling. 

In view of the significantly more complete eBOSS catalog available now
(see section labeled `Data' below), however, the BAO data at $z=2.334$ by themselves
should constitute an important constraint on the geometry of the Universe 
sampled solely by the quasars in this catalog (distributed at $z\gtrsim 1.77$)
and their Lyman-$\alpha$ forests. Since it is well known that a comparison
of the model predictions and the BAO measurements tends to become more
discordant with increasing redshift, it is desirable to use the Lyman-$\alpha$ 
data by themselves for model selection purposes, in parallel to the
already completed studies based on the BAO observations at various
redshifts.

In this {\sl Letter}, we therefore complement our previous model selection analysis 
by restricting the comparison to just the Lyman-$\alpha$ measurements. As
noted, we do this for two principal reasons. First, the model contrast at
$z\gtrsim 2$ is significantly greater than that at $z\sim 0$. Second, we now 
have the final BAO constraints from the Lyman-$\alpha$ auto-correlation and 
Lyman-$\alpha$-quasar cross-correlation functions produced from the sixteenth 
(DR16) and final release \cite{Ahumada:2020} of the fourth generation Sloan 
Digital Sky Survey (SDSS-IV), containing all of the clustering and 
Lyman-$\alpha$ data from the completed `extended' Baryonic Oscillation
Spectroscopic Survey (eBOSS) \cite{Dawson:2016}. This catalog of quasars
and Lyman-$\alpha$ profiles is so large (see `Data' below) that
the BAO scale measured from it at an effective redshift $\sim 2.334$ 
constitutes a crucial probe of the cosmology on its own merit. As we
shall see, these data alone provide an important comparison of cosmological 
models and their predictions without having to pre-assume $r_d$ and $H_0$. 
The latter is especially desirable in view of the growing disparity 
between the measurements of $H_0$ at low and high redshifts, creating a 
$\sim 4\sigma$ uncertainty in its value \cite{Riess:2021}. 

We shall find that the depth and extent of the final Lyman-$\alpha$ BAO analysis
\cite{duMas:2020}, along with the option of altogether avoiding the use of $r_d$ 
and $H_0$, provides us with a very clean and compelling test of the various
FLRW cosmologies. 

\section{Data}\label{data}
The analysis in this {\sl Letter} utilizes the baryonic acoustic oscillations (BAO) 
measured in the Lyman-$\alpha$ absorption and background quasars, with an effective 
(enemble) redshift $z=2.334$. The data are taken from the complete extended Baryonic 
Oscillation Spectroscopic Survey (eBOSS) \cite{duMas:2020}, which includes the 
Lyman-$\alpha$ absorption profiles of 210,005 background quasars distributed at 
$z_q>2.10$. The BAO scale has been measured in both the auto-correlation of the 
Lyman-$\alpha$ absorbers and their cross-correlation with 341,468 quasars at 
$z_q>1.77$. This data release represents several advances over previously 
published Lyman-$\alpha$ BAO measurements, including improved statistics from 
a larger quasar catalog and deeper observations, and a more accurate modeling of 
the systematics.

\section{Model Comparisons}\label{models}
The Lyman-$\alpha$ BAO survey measures the BAO scale in the Lyman-$\alpha$ forest
of absorption of light from distant quasars, both via the forest-forest correlation
function \cite{Delubac:2015}, and in the forest-quasar cross-correlation 
\cite{Font-Ribera:2014}. The scale is measured along the line-of-sight,
\begin{equation}
\Delta z = {r_d\over (1+z)\,d_H(z)}\;,\label{eq:Deltaz}
\end{equation}
and in the transverse direction,
\begin{equation}
\Delta\theta = {r_d\over d_M(z)}\label{eq:Deltatheta}
\end{equation}
where, as previously noted, $r_d$ is the length corresponding to the peak of the 
matter two-point function in comoving coordinates. In addition, the quantity
\begin{equation}
d_H(z)\equiv{c\over H(z)}\label{eq:dHz}
\end{equation}
is the Hubble scale at $z$, while $d_M(z)$ is given in terms of the angular-diameter
distance via the relation
\begin{equation}
d_M(z)\equiv (1+z)\,d_A(z)\;.\label{eq:dMz}
\end{equation}

\begin{table}
\begin{center}
\caption{Model comparison using the Lyman-$\alpha$ BAO data}\label{tab1}
\begin{tabular}{ll}
\hline\hline \\
Data/Model$\qquad\qquad$ & ${d_M/d_H}\quad\;\,$ ({\rm P-value})$\qquad\qquad$ \\ 
\\
\hline \\
Data                     & $4.17\pm0.18$ \\
\\
1. $R_{\rm h}=ct$           & $4.02\qquad\quad\;\;\;$ ($0.39$) \\ 
2. {\it Planck}-$\Lambda$CDM & $4.56\qquad\quad\;$ ($0.03$) \\ 
3. {\rm Milne}              & $5.058\quad$ ($<0.00001$) \\
4. {\rm Einstein-de Sitter} & $5.507\quad$ ($<0.00001$) \\ 
\\
\hline
\end{tabular}
\end{center}
\end{table}

We mention here that surveys often also define the distance
\begin{equation}
d_V(z)\equiv z^{1/3}d_M^{2/3}(z)\,d_H^{1/3}(z)\;,\label{eq:dV}
\end{equation}
an angle-weighted average of $d_M(z)$ and $d_H(z)$, but this quantity was 
not derived independently of $d_M$ and $d_H$ for these data \cite{duMas:2020}.
Thus, though it tends to be the best determined scale in measurements of BAO 
from galaxy surveys, we will not find it useful for our analysis in this particular 
instance, and we shall instead focus our attention on $d_M$ and $d_H$ themselves. 

Given that these BAO observables are proportional to $r_dH_0$ (see below for
the model-dependent functional form of $H[z]$), the ratios of the various BAO scales
are completely independent of the sonic horizon and Hubble constant. Indeed, as long
as we restrict our comparisons to these ratios, three of the models we compare
here have no free parameters at all, while the standard model, $\Lambda$CDM (see 
Eq.~\ref{eq:Ez} below), has just one: the scaled matter density $\Omega_{\rm m}$. 
But to avoid unduly `punishing' this cosmology by treating $\Omega_{\rm m}$ as an 
unknown variable when assessing its likelihood, we shall simply adopt its {\it Planck} 
optimized value, $\Omega_{\rm m}= 0.315\pm0.007$, with a spatial curvature constant
$k= 0$ \cite{PlanckVI:2020}. 

We shall compare fits to the Lyman-$\alpha$ data using four different 
cosmological models, each with its own prediction of the angular-diameter
and Hubble distances. The quantity $\Omega_i$ is the energy density 
of species $i$, scaled to the critical density
\begin{equation}
\rho_{\rm c}\equiv {3c^2H_0^2\over 8\pi G}\;,\label{eq:rhoc}
\end{equation}
in terms of the Hubble constant, $H_0=67.4\pm0.5$ km s$^{-1}$ Mpc$^{-1}$. In
principle, we could completely avoid any reliance on the {\it Planck} measurements 
by selecting the value of $\Omega_{\rm m}$ that optimizes $\Lambda$CDM's fit
to the Lyman-$\alpha$ BAO data, but the improvement is too small to justify the
introduction of an additional free parameter. The four models we compare are:

\begin{enumerate}

\item The $R_{\rm h}=ct$ universe, a Friedmann-Lema\^itre-Robertson-Walker cosmology
with zero active mass, $\rho+3p=0$, in terms of the total energy density ($\rho$)
and pressure ($p$) in the cosmic fluid \cite{MeliaShevchuk:2012,Melia:2020}. In this case,
\begin{equation}
d^{\,(1)}_M(z)=\frac{c}{H_0}\ln(1+z)\;,\label{eq:dMRh}
\end{equation}
and
\begin{equation}
d^{\,(1)}_H(z) = \frac{c}{H_0(1+z)}\;.\label{eq:dHRh}
\end{equation}
\vskip 0.1in

\item Flat {\it Planck}-$\Lambda$CDM, with $\Omega_\Lambda=1-\Omega_{\rm m}$ and a
dark-energy equation of state parameter $w_{\rm de}=-1$. For this model,
\begin{equation}
d^{\,(2)}_M(z)=\frac{c}{H_0}
\int _0^{\,z}\frac{du}{E(u)}\;,\label{eq:dMLCDM}
\end{equation}
and
\begin{equation}
d^{\,(2)}_H(z) = \frac{c}{H_0E(z)}\;,\label{eq:dHLCDM}
\end{equation}
where
\begin{equation}
E(z) \equiv \left[\Omega_{\rm m}(1+z)^3+\Omega_\Lambda\right]^{1/2}\;.\label{eq:Ez}
\end{equation}
At these redshifts, the contribution to $E(z)$ from radiation is negligible, so
we have ignored it for the calculation of $d_H(z)$ and $d_M(z)$.
\vskip0.1in

\item The Milne universe. This well studied solution is also a Friedmann-Lema\^itre-Robertson-Walker
cosmology, but its energy density, pressure and cosmological constant are all zero. Its expansion
instead derives from a non-zero spatial curvature, with $k=-1$. Like the $R_{\rm h}=ct$
universe, the Milne scale factor, $a(t)$, is also linear in time 
\cite{Vishwakarma:2013,Chashchina:2015}, but we include it here primarily because the
observable signatures in these two models are very different. In the Milne universe,
\begin{equation}
d^{\,(3)}_M(z)=\frac{c}{H_0}\sinh\left[\ln(1+z)\right]\;,\label{eq:dMMilne}
\end{equation}
and
\begin{equation}
d^{\,(3)}_H(z) = \frac{c}{H_0(1+z)}\;.\label{eq:dHMilne}
\end{equation}

\item Einstein-de Sitter (i.e., Eqs.~\ref{eq:dMLCDM} and \ref{eq:dHLCDM} with
$\Omega_{\rm m}=1$ and $\Omega_\Lambda=0$):
\begin{equation}
d^{\,(4)}_M(z)=\frac{2c}{H_0}\left(1-\frac{1}{\sqrt{1+z}}\right)\;,\label{eq:dMEdS}
\end{equation}
and
\begin{equation}
d^{\,(4)}_H(z) = \frac{c}{H_0(1+z)^{3/2}}\;.\label{eq:dHEdS}
\end{equation}
This cosmology is already heavily disfavoured by many other observations, but we
include it here because of its relevance as a former `standard' model. 
\vskip 0.1in

\end{enumerate}

The combined BAO measurements from the auto- and cross-correlation in the final, complete
eBOSS release \cite{duMas:2020} yield the following constraints:
\begin{eqnarray}
d_H(z=2.334)/r_d &=& 8.99\pm0.19\nonumber \\
d_M(z=2.334)/r_d &=& 37.5\pm1.1\;.\label{eq:BAOfinal}
\end{eqnarray}
Though a `fiducial' cosmology is employed in the calculation of these
distances, the ratios shown in Equation~(\ref{eq:BAOfinal}) are model
independent, as studied in detail by ref.~\cite{Carter:2020} in the context
of galaxy correlations, and confirmed for the Lyman-$\alpha$ observations
by ref.~\cite{duMas:2020}.

The model predictions are compared with these data in Table~1, prioritized
in terms of the p-values estimated from the various fits. Given
a model's prediction for $R^{\rm th}\equiv d_M/d_H$ and the standard deviation 
(in this case, $\sigma_R=0.18$) of the measurement, this p-value represents the 
probability of observing a difference $|R^{\rm obs}-R^{\rm th}|$ greater than
$|4.17-R^{\rm th}|$ under the assumption that the null hypothesis is true
and that the distribution of $R^{\rm obs}$ is normal. The error quoted
for the measured value of $d_M/d_H$ includes the correlation between
$d_M/r_d$ and $d_H/r_d$, characterized by the correlation coefficient
$C(d_H,d_M)=-0.45$ \cite{duMas:2020}. Small p-values provide
evidence against the assumed model, with the evidence getting stronger
as the p-value approaches zero. Typical guidelines suggest the following
hierarchy: ($p>0.10$) weak or no evidence; ($0.05<p\le 0.10$) moderate
evidence; ($0.01<p\le 0.05$) strong evidence; ($p\le 0.01$) very strong
evidence.

On the basis of these comparisons, it is clear that the Lyman-$\alpha$
BAO measurements very strongly rule out the Milne and Einstein-de Sitter
cosmologies, affirming an already established conclusion drawn from many
other comparative tests \cite{Melia:2020}. There is also strong evidence
disfavoring the standard model, an outcome confirmed by several previous 
studies, including those reported in refs.~\cite{duMas:2020} and \cite{Evslin:2017}.
But the principal result of this work is that the Lyman-$\alpha$ BAO
measurements are completely consistent with the geometry of the cosmos
predicted by the $R_{\rm h}=ct$ universe.

Pursuing this head-to-head comparison further, we note that if $\Omega_{\rm m}$
were to differ from its {\it Planck} value by $3\sigma$, i.e., if 
$\Omega_{\rm m}=0.294$, then the p-value for $\Lambda$CDM would
improve somewhat to $0.065$, so the evidence against the standard
model in that case would be `moderate' instead of `strong.'
Nevertheless, to make the standard model equally likely with
$R_{\rm h}=ct$, i.e., to improve its p-value to $0.39$, $\Omega_{\rm m}$
would have to be smaller than $\sim 0.236$, a value different by many
standard deviations from that inferred by {\it Planck}.

It is also straightforward to perform a Bayesian analysis 
\cite{Dienes:2016,Kruschke:2017} of the head-to-head comparison 
between $\Lambda$CDM and $R_{\rm h}=ct$ based on the Lyman-$\alpha$ 
data at $z=2.334$, specifically, the measurement of $d_M/d_H=4.17\pm0.18$ 
shown in Table 1. Under the assumption that this ratio follows a normal 
distribution, and adopting a point model for both the null and alternative 
hypotheses (remember that we are fixing the value of $\Omega_{\rm m}$ 
at the {\it Planck} measurement to avoid unduly `punishing' the standard 
model for having an additional free parameter), one infers a Bayes factor 
of 7.39. The marginal likelihoods at the measured value of $d_M/d_H$,
from which the Bayes factor is estimated, are shown in Figure~1. 

\begin{figure}
\begin{center}
\includegraphics[width=0.8\linewidth]{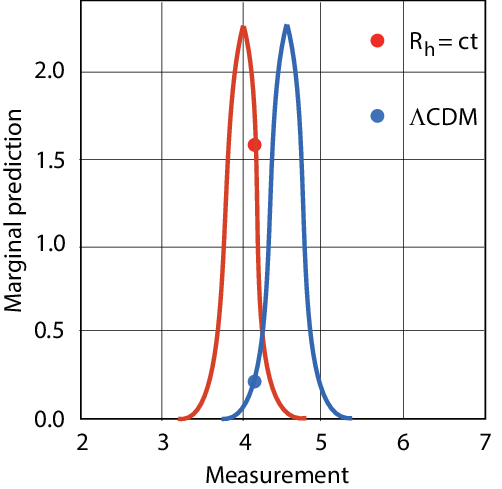}
\end{center}
\caption{Estimation of the Bayes factor from the marginal prediction
of the $R_{\rm h}=ct$ and $\Lambda$CDM models versus the measurement
of the ratio $d_M/d_H$ from the Lyman-$\alpha$ forest at an effective
redshift $z=2.334$.}
\end{figure}

A Bayes factor between 3 and 10 (as we have here) indicates `moderate' 
or `substantial' evidence in favor of the alternative hypothesis
\cite{Jeffreys:1961,Kass:1995,Wetzels:2011}. As such, the Bayesian
analysis complements and confirms our earlier conclusion, derived
from the p-values, that the Lyman-$\alpha$ BAO data at $z=2.334$
favor the $R_{\rm h}=ct$ cosmology over the current standard model.

\section{Conclusion}\label{conclusion}
The strong rejection of the Milne and Einstein-de Sitter cosmologies
by the final Lyman-$\alpha$ BAO data release is hardly surprising
in view of their similarly strong rejection by other observations.
Our main conclusion from this study instead refocuses our attention
on the fact that the data tend to favour the $R_{\rm h}=ct$ cosmology
over the current standard model. 

Indeed, the BAO measurements offer a new perspective on this comparison,
following an earlier examination by ref.~\cite{Evslin:2017}, who attempted
to identify the reason behind the BAO anomaly in the context of $\Lambda$CDM.
Their analysis showed that the BAO data at $z>0.43$ are in tension with
the standard model, whether or not the {\it Planck} optimized parameters
(e.g., for $\Omega_{\rm m}$) are assumed. They concluded that this tension
arises not from the $\Lambda$CDM parameters, but instead from the
dark energy evolution at $0.57<z<2.334$. If one further sets $r_d$
equal to the acoustic scale measured in the CMB, a cosmological constant
for dark energy is firmly rejected. 

The $R_{\rm h}=ct$ cosmology is essentially $\Lambda$CDM, though with the
crucial additional constraint of zero active mass from general relativity
\cite{Melia:2020}. This constraint features an equation-of-state,
$\rho+3p=0$, in terms of the total energy density $\rho$ and pressure $p$
in the cosmic fluid. Dark energy is therefore dynamic in this model, 
evolving along with the other constituents, presumably an extension to 
the standard model of particle physics. 

The current standard model suffers from several major conflicts and 
inconsistencies that continue to defy serious attempts at mitigation
\cite{Melia:2022e}. Together with the continued success of $R_{\rm h}=ct$
to account for the data better than $\Lambda$CDM \cite{Melia:2018e},
as affirmed by the work reported in this {\sl Letter}, the prospects for further
development of this alternative FLRW cosmology look very promising.
An especially exciting future observation to anticipate over the coming 
decade is the real-time measurement of redshift drift \cite{Melia:2016b,Melia:2022d},
which should provide an unambiguous yes/no answer to the question of whether
or not the cosmic fluid is in fact driven by a zero active mass equation-of-state. 

%\acknowledgments

\bibliographystyle{eplbib}
\bibliography{ms}

\end{document}